\documentclass[prl,twocolumn,floatfix,superscriptaddress]{revtex4}
\usepackage{graphicx}

\newlength{\onecolfig}
\newlength{\twocolfig}
\newcommand{\ion}[2]{\mbox{$^{#2}$#1$^+$}}
\newcommand{\Ca}[1]{\ion{Ca}{#1}}

\newcommand{\lev}[2]{\mbox{#1$_{\mbox{\tiny$#2$}}$}}
\newcommand{\hfslev}[3]{\mbox{#1$^{\mbox{\tiny$#3$}}_{\mbox{\tiny$#2$}}$}}

\newcommand{\unit}[1]{\,\mbox{#1}}
\newcommand{\mHz}{\unit{mHz}}
\newcommand{\Hz}{\unit{Hz}}
\newcommand{\kHz}{\unit{kHz}}
\newcommand{\MHz}{\unit{MHz}}
\newcommand{\GHz}{\unit{GHz}}

\newcommand{\torr}{\unit{torr}}

\newcommand{\mW}{\unit{mW}}
\newcommand{\uW}{\unit{$\mu$W}}

\newcommand{\um}{\unit{$\mu$m}}
\newcommand{\nm}{\unit{nm}}

\newcommand{\s}{\unit{s}}
\renewcommand{\sec}{\unit{sec}}

\newcommand{\persec}{\unit{s$^{-1}$}}
\newcommand{\ms}{\unit{ms}}
\newcommand{\us}{\unit{$\mu$s}}

\newcommand{\degree}{\mbox{$^{\circ}$}}

\newcommand{\G}{\unit{G}}
\newcommand{\mG}{\unit{mG}}
\newcommand{\dB}{\unit{dB}}

\newcommand{\citesec}[2]{\cite[\S{}#2]{#1}}   

\newcommand{\eg}{{\em e.g.}}

\newcommand{\ish}{\mbox{$\sim$}\,}
\newcommand{\ltish}{\protect\raisebox{-0.4ex}{$\,\stackrel{<}{\scriptstyle\sim}\,$}}

\newcommand{\ket}[1]{\mbox{$\left| #1 \right>$}}

\newcommand{\sub}[1]{\mbox{$_{\mbox{\tiny #1}}$}}
\newcommand{\diff}[1]{\mbox{\/d$#1$}}

\newenvironment{centre}{\begin{center}}{\end{center}}

\begin{document}

\title{High-fidelity preparation, gates, memory and readout of a trapped-ion quantum bit}

\author{T. P. Harty}
\affiliation{Department of Physics, University of Oxford, Clarendon Laboratory, Parks Road, Oxford OX1 3PU, U.K.}
\author{D. T. C. Allcock}
\affiliation{Department of Physics, University of Oxford, Clarendon Laboratory, Parks Road, Oxford OX1 3PU, U.K.}
\author{C. J. Ballance}
\affiliation{Department of Physics, University of Oxford, Clarendon Laboratory, Parks Road, Oxford OX1 3PU, U.K.}
\author{L. Guidoni}
\affiliation{Department of Physics, University of Oxford, Clarendon Laboratory, Parks Road, Oxford OX1 3PU, U.K.}
\affiliation{University of Paris Diderot, Laboratoire Mat\'eriaux et Ph\'enom\`enes Quantiques, UMR 7162 CNRS, F-75205 Paris, France}
\author{H. A. Janacek}
\affiliation{Department of Physics, University of Oxford, Clarendon Laboratory, Parks Road, Oxford OX1 3PU, U.K.}
\author{N. M. Linke}
\affiliation{Department of Physics, University of Oxford, Clarendon Laboratory, Parks Road, Oxford OX1 3PU, U.K.}
\author{D. N. Stacey}
\affiliation{Department of Physics, University of Oxford, Clarendon Laboratory, Parks Road, Oxford OX1 3PU, U.K.}
\author{D. M. Lucas} 
\affiliation{Department of Physics, University of Oxford, Clarendon Laboratory, Parks Road, Oxford OX1 3PU, U.K.}
\email{d.lucas@physics.ox.ac.uk}

\date{7 Oct 2014, v9arxiv}


\begin{abstract}
We implement all single-qubit operations with fidelities significantly above the minimum threshold required for fault-tolerant quantum computing, using a trapped-ion qubit stored in hyperfine ``atomic clock'' states of \Ca{43}. We measure a combined qubit state preparation and single-shot readout fidelity of 99.93\%, a memory coherence time of {$T^*_2=50$}~seconds, and an average single-qubit gate fidelity of 99.9999\%. These results are achieved in a room-temperature microfabricated surface trap, without the use of magnetic field shielding or dynamic decoupling techniques to overcome technical noise. 
\end{abstract}

\maketitle



The great potential of quantum computing requires two essential ingredients for its realization: high-fidelity quantum logic operations and a physical implementation which can be scaled up to large numbers of quantum bits~\cite{Nielsen00}. Amongst the candidate technologies for implementing quantum information processing, individual trapped ions were recognized early as a very promising system~\cite{Cirac95,Steane97,Wineland98}: the qubits are stored in internal atomic energy levels of the ions, which can be extremely stable and well isolated from the environment, and the strong Coulomb interaction between neighbouring ions can be used to mediate qubit-qubit logic. Since the first proposals, multiple-qubit algorithms have been demonstrated~\cite{Blatt08}, and there has been significant progress in developing scalable ion trap technologies~\cite{Monroe13}. Long qubit memory coherence time~\cite{Langer05}, high-fidelity state preparation and readout~\cite{Myerson08}, and single-qubit gates with fault-tolerant error rates~\cite{Brown11} have all been demonstrated, in a variety of different trapped ions and experiments. 

In this Letter, we demonstrate all single-qubit operations (preparation, memory, gates and readout) with performances comparable to or better than previous work, and all in the same system. All errors are more than an order of magnitude below the $\approx 1\%$ fault-tolerant thresholds emerging from recent numerical calculations using surface-code error correction~\cite{Fowler12b}; this is critical for the practical implementation of fault-tolerant methods, whose resource requirements increase dramatically for error rates close to threshold~\cite{Steane03}. Furthermore, the ion-qubit is trapped in a microfabricated surface-electrode trap~\cite{Seidelin06} with a two-dimensional electrode layout which is extendable to large arrays of multiplexed traps, as envisaged in the original proposal for scalable trapped-ion quantum information processing~\cite{Wineland98}. We describe below the trap and the \Ca{43} qubit, and three experiments performed to measure the combined state preparation and readout error, the qubit coherence time, and the average single-qubit gate error.

\begin{figure*}
\includegraphics[width=\twocolfig]{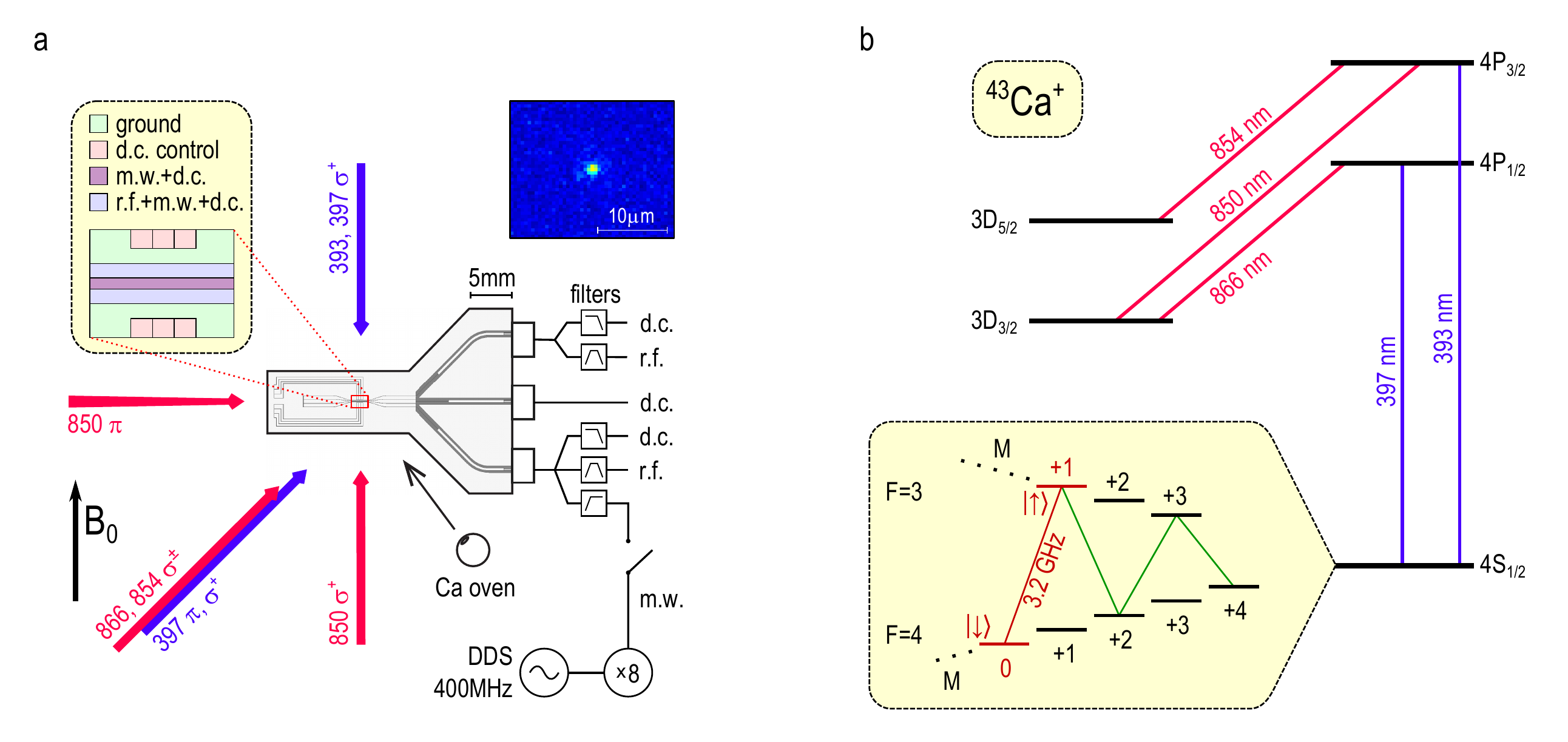}
\caption{%
The ion trap and the qubit. (a) Schematic diagram of the surface ion trap, showing (left inset) central electrode layout. Microwave (m.w., 3.2\GHz) signals are combined with the trap radiofrequency voltage (r.f., 40\MHz) via filters, as indicated here for the lower axial electrode. Also shown are laser beam directions and polarizations with respect to the static magnetic field $B_0=146\G$. The violet 397\nm\ Doppler-cooling beam is elliptically polarized such that it contains only $\pi$ and $\sigma^+$ polarizations, and co-propagates with linearly ($\sigma^\pm$) polarized infra-red repumping beams. Circularly $\sigma^+$ and linearly $\pi$ polarized beams are used for state preparation and readout (the $\pi$ beam is at $\ish 45\degree$ to the plane of the figure and reflects off the trap surface). To load the trap, neutral Ca atoms effusing from the oven are ionized by laser beams at 423\nm\ and 389\nm\ which co-propagate with the Doppler-cooling beams. Ion fluorescence is collected by an imaging system perpendicular to the plane of the trap; the right inset shows an image of a single \Ca{43} ion, which is trapped 75\um\ above the electrodes. %
(b) \Ca{43} level structure, showing optical transitions used for Doppler cooling, qubit state preparation and readout. The inset shows part of the ground level hyperfine structure, labelled by quantum numbers $F$ and $M$, with the qubit \ket{\downarrow} and \ket{\uparrow} states, the 3.2\GHz\ qubit transition (red) and the auxiliary transitions (green) used for state preparation and readout. The Zeeman splittings between adjacent $M$ states are $\approx 50\MHz$.
}
\label{F:trapqubit}
\end{figure*}

The ion trap is of a novel design which incorporates integrated microwave circuitry (resonators, waveguides, and coupling elements), designed to allow single- and two-qubit quantum logic gates to be driven by near-field microwaves~\cite{Ospelkaus11, Warring13a} instead of by lasers: this will enable all the coherent qubit operations to be performed by electronic techniques, where one can take advantage of readily available microwave sources whose power and absolute frequency are very stable, and which can be easily connected to the trap electrodes. In contrast to solid-state qubit technologies~\cite{Devoret13}, it is not necessary to cool the apparatus to milli-Kelvin temperatures, as the microwave control fields are classical: only the qubits themselves need to be cold, and this is straightforwardly achieved using Doppler laser-cooling. A schematic diagram of the trap and the laser beam layout is shown in fig.~\ref{F:trapqubit}a; the trap is described in more detail in ref.~\cite{Allcock13}.  

A single \Ca{43} ion is loaded into the trap from a 12\%-enriched calcium source using isotope-selective photo-ionization~\cite{Lucas04}. Trap secular frequencies are typically 3\MHz\ (radial) and 500\kHz\ (axial). The ion is Doppler-cooled with lasers operating at 397\nm\ and 866\nm; further lasers at 393\nm, 850\nm\ and 854\nm\ are used for qubit readout and reset. An advantage of the \Ca{} ion is that all wavelengths are available from solid-state diode lasers without the need for frequency-doubling, are compatible with integrated optics~\cite{Brady11}, and do not cause observable charging of the trap structure under normal operation. The optical operations (cooling, state preparation and readout) are robust to laser intensity and frequency noise (laser linewidths are $\approx 1\MHz$), and only require low-power beams. 

\begin{figure}
\includegraphics[width=\onecolfig]{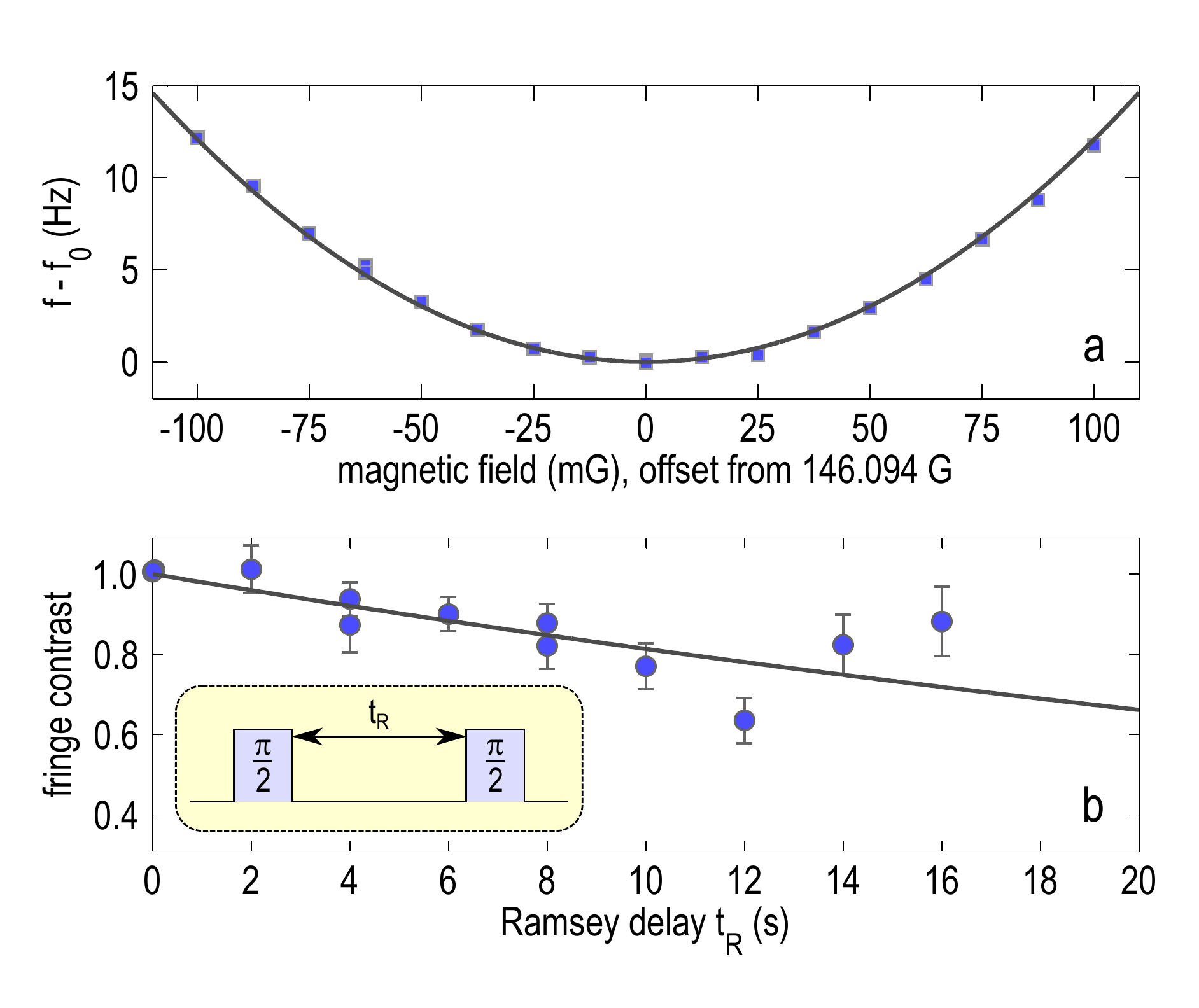}
\caption{%
(a) Microwave spectroscopy of the qubit transition, varying the static magnetic field $B$ through the field-independent point $B_0=146.094\G$. At each field value, the qubit transition frequency $f$ was measured by Ramsey spectroscopy, to a precision $\approx 0.1\Hz$. The field-independent qubit transition is at $f_0=3\,199\,941\,077\Hz$ after adjusting for a $-5\Hz$ shift due to the trap (see text). The solid line shows the expected frequency calculated using the Breit-Rabi formula assuming the known zero-field hyperfine splitting~\cite{Arbes94} and a nuclear magnetic moment~\cite{TPHthesis} of $\mu_I=-1.31535\mu_N$. (b) Qubit coherence time measurements. At each value of the Ramsey free precession time $t_R$ the phase of the second $\pi/2$ pulse was varied to produce a set of Ramsey fringes. The contrast of the fringes is fitted with an exponential decay, giving a coherence time $T^*_2=50(10)\s$.
}
\label{F:clock}
\label{F:Ramsey}
\end{figure}

Hyperfine states in the ground 4\lev{S}{1/2} level of the ion are used for the qubit states (fig.~\ref{F:trapqubit}b). As spontaneous decay rates are negligible, these states have essentially infinite $T_1$ times (limited in practice by the ion trapping lifetime, which is typically several hours in this trap under ultra-high vacuum conditions, $<10^{-11}\torr$). $T_2$ coherence times are limited by the frequency stability of the qubit transition. The state energies depend on the static magnetic field $B$ through the Zeeman effect and ambient magnetic field noise would normally limit the coherence time to a few ms. However, certain transition energies become independent of magnetic field to first order at particular values of the field, due to the non-linear dependence arising from hyperfine state mixing, and these permit particularly stable qubits~\cite{Langer05}. We choose one of these so-called ``atomic clock'' transitions, $\hfslev{S}{1/2}{4,0}\leftrightarrow\hfslev{S}{1/2}{3,+1}$ (where the superscripts denote angular momentum quantum numbers $F, M$), which in \Ca{43} is field-independent at $B_0\approx 146\G$ (fig.~\ref{F:clock}a). Unlike schemes for microwave quantum logic based on static magnetic field gradients~\cite{Mintert01}, the use of near-field microwaves allows all qubits to share the same, well-defined, noise-immune, frequency.

The relatively large magnetic field leads to a complex atomic level structure, with Zeeman splittings spanning $\ish 500\MHz$, and because of the low-lying D levels in \Ca{} there is no closed cycling transition for laser cooling. We have nevertheless identified a simple Doppler cooling method which requires only two 397\nm\ frequencies, a single 866\nm\ frequency, moderate laser powers ($\ish 100\uW$) and a single beam direction. We obtain a fluorescence count rate comparable to that from a single, saturated, \Ca{40} ion, at $50\,000\persec$ with a net photon detection efficiency of 0.3\%, which is sufficient for high-fidelity fluorescence detection. 


To measure the combined state preparation and measurement (SPAM) error we repeatedly prepare the same qubit state, and read it out, averaging over preparations of the \ket{\downarrow} and \ket{\uparrow} states. We first optically pump the ion to the \hfslev{S}{1/2}{4,+4} state using circularly $\sigma^{+}$ polarized 397\nm\ light. We use a microwave technique to improve the optical pumping fidelity, which is otherwise limited by imperfect polarization of the $397\nm$ beam (see Supplementary Information). A series of three (or four) microwave $\pi$-pulses on the transitions indicated in the inset to fig.~\ref{F:trapqubit}b then transfers the ion to the \ket{\uparrow} (or \ket{\downarrow}) qubit state, as desired. To read out the qubit state, three microwave $\pi$-pulses transfer population in \ket{\uparrow} to \hfslev{S}{1/2}{4,+4}, and a fourth $\pi$-pulse transfers $\ket{\downarrow}\rightarrow\hfslev{S}{1/2}{3,+1}$. Population in \hfslev{S}{1/2}{4,+4} is then ``shelved'' in the metastable 3\lev{D}{5/2} level by a repeated sequence of ($393\nm\ \sigma^+, 850\nm\ \sigma^+, 850\nm\ \pi$-polarized) optical pumping pulses, as described in~\cite{Myerson08}. Finally the Doppler-cooling lasers are applied again and we detect whether or not the ion was shelved by the absence or presence of 397\nm\ fluorescence. 

\begin{figure}
\includegraphics[width=\onecolfig]{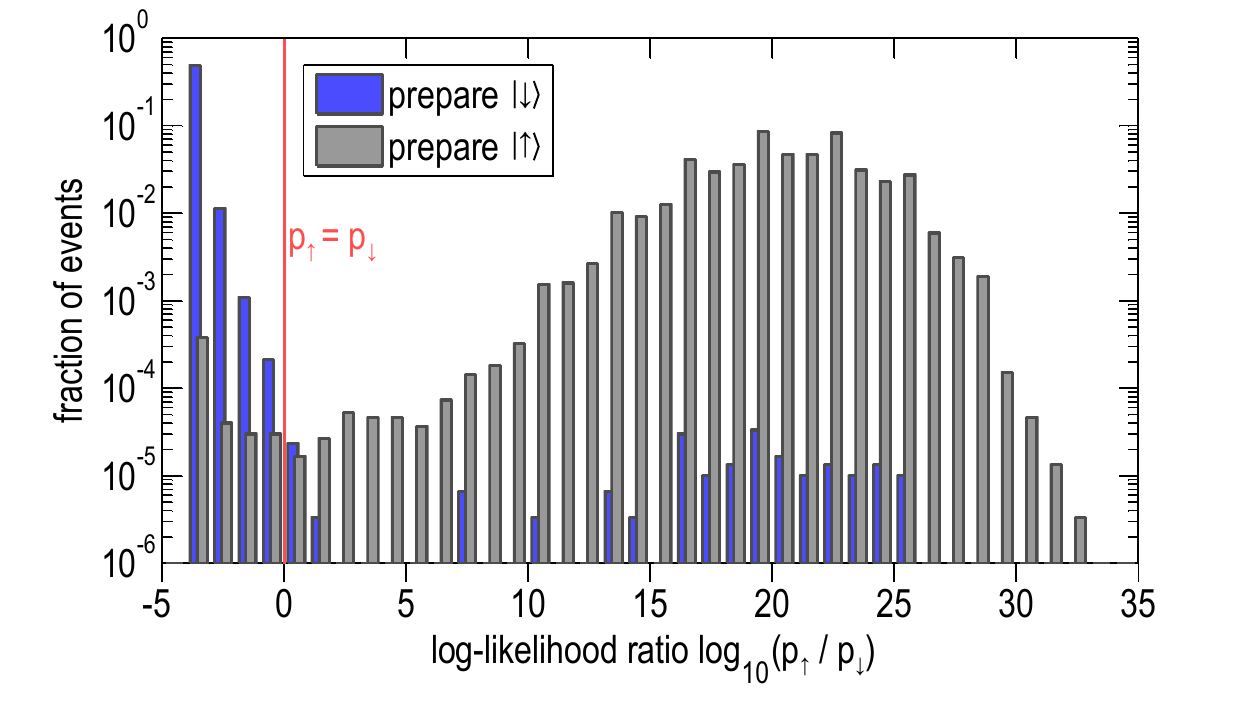}
\caption{%
Qubit state preparation and measurement (SPAM) results. We prepared and measured the $\ket{\downarrow}$ (blue histogram) and $\ket{\uparrow}$ (grey histogram) qubit states $150\,000$ times each. For each measurement the likelihood $p_\downarrow$ (or $p_\uparrow$) that the state was $\ket{\downarrow}$ (or $\ket{\uparrow}$) was calculated from the time-resolved photon counts detected on a photomultiplier (see~\cite{Myerson08}); if $p_\uparrow>p_\downarrow$ we infer that the qubit was in the $\ket{\uparrow}$ state, and vice versa. If the state inferred disagrees with the state prepared, either a preparation or a measurement error has occurred; thus the fraction of experiments in which an error occurred is given by the sum of the blue events above the $p_\uparrow=p_\downarrow$ threshold ($2\times 10^{-4}$) and the grey events below threshold ($5\times 10^{-4}$). 
}
\label{F:readout}
\end{figure}

For $150\,000$ preparations of each qubit state, we measure the combined SPAM error to be $6.8(5)\times 10^{-4}$ (fig.~\ref{F:readout}). As the qubit readout method is not a quantum non-demolition measurement, we cannot repeat it many times to separate the preparation and readout errors, but from estimates of the various contributions to the combined error (table~\ref{T:errors}) we assign errors of $\approx 2\times 10^{-4}$ to the state preparation and $\approx 5\times 10^{-4}$ to the readout. The error contributions could all be reduced by technical improvements (e.g.\ increasing the photon detection efficiency~\cite{Myerson08}), except for that due to the shelving transfer to \lev{D}{5/2} which is limited to a minimum~\cite{TPHthesis} of $\approx 1\times 10^{-4}$ (at $B_0=146\G$) by the atomic structure of \Ca{43}. 

The qubit coherence time was measured by performing Ramsey experiments (without any dynamic decoupling pulses~\cite{Szwer11}) on the $\hfslev{S}{1/2}{4,0}\leftrightarrow\hfslev{S}{1/2}{3,+1}$ qubit transition at $f_0=3.200\GHz$. To ensure that the applied magnetic field remained close to the field-independent point, the frequency of the field-{\em dependent} $\hfslev{S}{1/2}{4,+4}\leftrightarrow\hfslev{S}{1/2}{3,+3}$ transition was periodically measured by the computer controlling the experiment, and an appropriate correction was applied to the magnetic field coil current. Ramsey delays up to $t_R = 16\sec$ were used, with results shown in fig.~\ref{F:Ramsey}b. An exponential decay $\exp(-t_R/T^*_2)$ fitted to the data gives a coherence time $T^*_2=50(10)\sec$. The coherence time may be limited by residual magnetic field drift (the qubit's second-order field dependence is $\diff{^2 f}/\diff{B^2} = 2.4\mHz/\mG^2$), instability of the local oscillator, and fluctuations in the amplitude of the trap r.f.\ voltage (by varying the r.f.\ power and extrapolating to zero power, we measure a differential a.c.\ Zeeman shift of $-5\Hz$ due to r.f.\ currents in the trap electrodes~\cite{Berkeland98a}). The reduction in fringe contrast could also be due to effects unrelated to the qubit coherence, for example heating of the ion during $t_R$ which increases readout error due to Doppler-broadening of the 393\nm\ shelving transition. We note that longer coherence times have been measured in large ensembles, using trapped ions~\cite{Bollinger91} and nuclear spins~\cite{Saeedi13} (in the latter case, only with multiple dynamical decoupling pulses). 


The fidelity of single-qubit gates driven by one of the near-field integrated microwave electrodes was measured by the established technique of randomized benchmarking~\cite{Knill08}, which yields an average gate error appropriate to a computational context. We use the same method as ref.\cite{Brown11}, which reports the previous lowest single-qubit gate error. Having prepared the qubit in \ket{\uparrow}, we apply a pre-programmed pseudo-random sequence of logical gates, where each logical gate comprises a Pauli gate followed by a Clifford gate. The sequence terminates by rotating the qubit into either \ket{\downarrow} or \ket{\uparrow}, chosen with equal probability. Clifford gates are randomly chosen to rotate the qubit about the $\pm x$ or $\pm y$ axes on the Bloch sphere; Pauli gates are randomly chosen to rotate about the $\pm x$, $\pm y$ or $\pm z$ axes, or to be a $\pm I$ identity gate. In the experiment, each Clifford gate is performed by a microwave $\pi/2$-pulse and each Pauli gate by a pair of $\pi/2$-pulses. Identity gates are implemented using delays of the same duration (12\us) as the $\pi/2$-pulses, $\pm z$ rotations as an identity followed by a rotation of the logical frame of the qubit for subsequent pulses. The microwaves are generated by a frequency-octupled 400\MHz\ direct digital synthesis (DDS) source, fed via a switch to one of the m.w.\ electrodes (fig.~\ref{F:trapqubit}a); the enhancement provided by the integrated m.w.\ resonator and the proximity of the ion to the electrode means that a low m.w.\ power (0.1\mW) is sufficient and a power amplifier is not necessary. The m.w.\ power was periodically calibrated during the experiments using a sequence of 751 $\pi/2$-pulses. The qubit was kept at the field-independent point by servoing the magnetic field as in the coherence time measurements.

Each pseudo-random sequence is applied many times, and we compare the measured final qubit state with the expected outcome for that sequence. We apply sequences of various lengths, up to 2000 computational gates, and use 32 distinct sequences at each length, for a total of 224 randomly-chosen sequences. Since some sequences are more susceptible to errors than others, we performed numerical simulations to check that this provided sufficient randomization over systematic variations (see Supplementary Information). Results are shown in fig.~\ref{F:gate}, where we deduce an average error per gate of $1.0(3)\times 10^{-6}$ from the slope of the fitted line. This is an upper limit since it ignores any possible increase in SPAM error for the longer sequences; an independent experiment comparing 2000-gate runs with control runs which contained no gates, but had the same 160\ms\ delay between preparation and readout, gave an error per gate of $0.4(8)\times 10^{-6}$. Estimated contributions to the measured gate error are shown in table~\ref{T:errors}; these can all be reduced or compensated for by technical improvements (for example, a trap design allowing arbitrary control of the microwave polarization~\cite{Aude14} could eliminate the off-resonant excitation of other m.w.\ transitions). 

\begin{figure}
\includegraphics[width=\onecolfig]{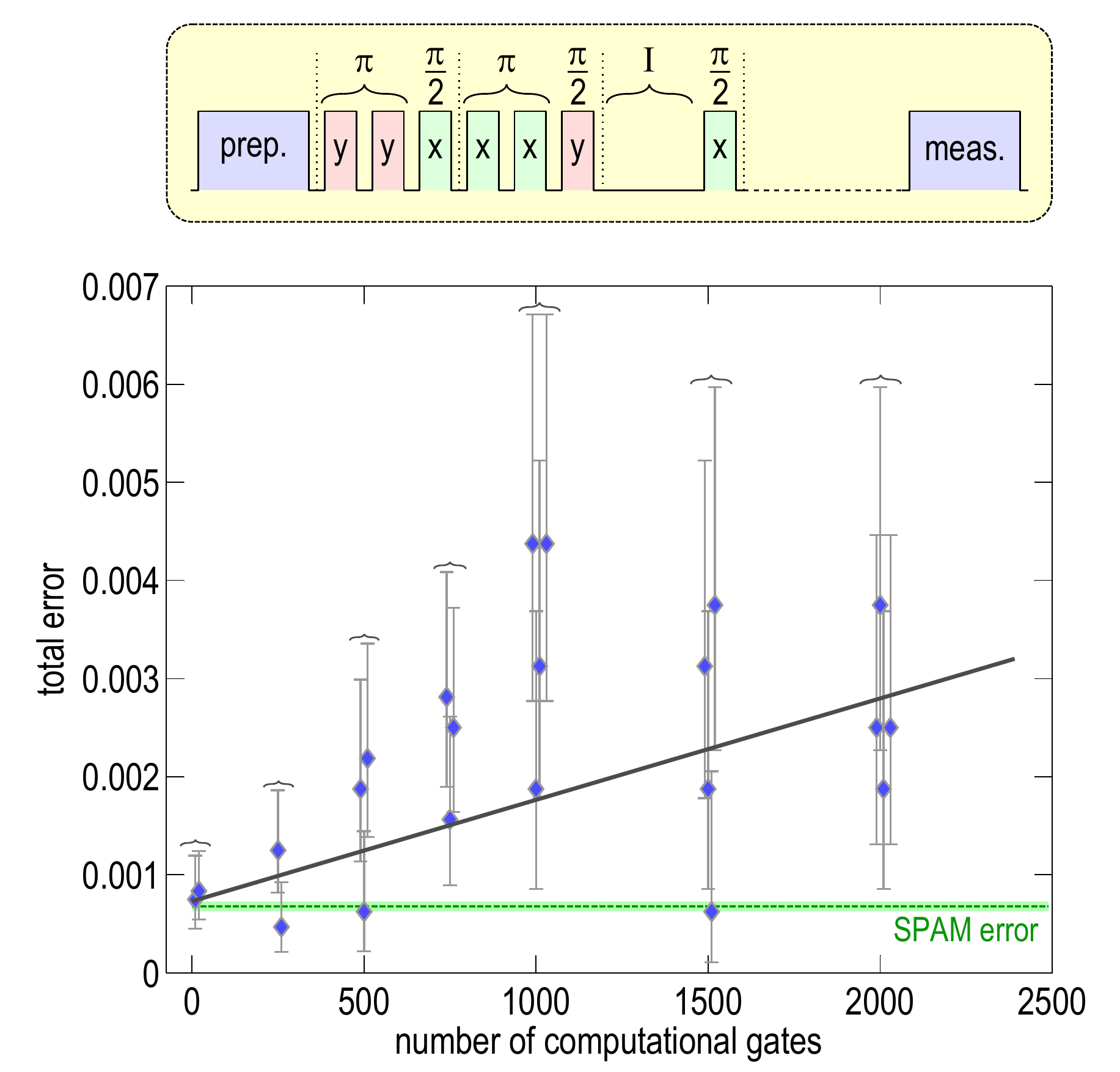}
\caption{%
Randomized benchmarking of single-qubit gates. The start of an example pseudo-random sequence is shown: each computational gate is comprised of three physical $\pi/2$ pulses (absent in the case of identity $I$ operations or $z$ rotations). Additional $\pi/2$ pulses at the end of the sequence rotate the qubit into the (\ket{\downarrow},\ket{\uparrow}) basis for measurement. If the measured state disagrees with that expected, an error is recorded. For each sequence length, 32 distinct sequences are used, each one being repeated the same number of times (typically 100) to measure the total error. Results are shown in the plot, where repeated runs have been offset horizontally for clarity. The gradient of a weighted straight line fit gives an average error per gate of $1.0(3)\times 10^{-6}$, while the intercept agrees with the independently measured SPAM error of $6.8(5)\times 10^{-4}$ (dashed line). Error bars represent 68\% confidence intervals, assuming binomial statistics. The fit gives a reduced $\chi^2 \approx 0.84$.
}
\label{F:gate}
\end{figure}

\begin{table}
\begin{centre}
\begin{tabular}{|l|c|} \hline
{\bf preparation/readout operation}                                      &  {\bf error}              \\ \hline
stretch state \hfslev{S}{1/2}{4,+4} preparation                          & $ < 1   \times 10^{-4} $  \\
transfer to qubit (3 or 4 m.w.\ $\pi$-pulses)                            & $ 1.8     \times 10^{-4} $  \\
transfer from qubit (4 m.w.\ $\pi$-pulses)                               & $ 1.8     \times 10^{-4} $  \\
shelving transfer $\hfslev{S}{1/2}{4,+4}\rightarrow\lev{D}{5/2}$         & $ 1.7   \times 10^{-4} $  \\
time-resolved fluorescence detection                                     & $ 1.5   \times 10^{-4} $  \\ \hline \hline
{\bf single-qubit gate error source}                                     &  {\bf mean EPG}           \\ \hline
microwave detuning ($4.5\Hz$)                                            & $ 0.7   \times 10^{-6} $  \\
microwave pulse area ($5\times 10^{-4}$)                                 & $ 0.3   \times 10^{-6} $  \\  
off-resonant effects                                                     & $ 0.1   \times 10^{-6} $  \\ \hline
\end{tabular}
\end{centre}
\caption{%
Error contributions: (top) state preparation and readout experiment; (bottom) single-qubit randomized benchmarking experiment (EPG: error per gate). The error contributions are estimates based on auxiliary experiments, experimentally measured parameters, and theoretical models of the various processes (see Supplementary Information and ref.~\cite{DJSthesis}).
}
\label{T:errors}
\end{table}


In conclusion, we have used a new magnetic-field-independent qubit in \Ca{43}, held in a scalable ion trap design, to demonstrate all single-qubit operations at error rates more than an order of magnitude below the threshold necessary for surface-code quantum error correction. The coherence time and gate fidelity surpass measurements in all other single physical qubits. The combined state preparation and readout error is the lowest measured for any ``atomic clock'' qubit. Although we did not need to employ composite pulse techniques to correct for technical noise, the exceedingly low single-qubit gate error means that the extensive library of such techniques~\cite{Levit86} is usable with negligible error overhead. In separate experiments~\cite{Ballance14}, we have demonstrated laser-driven two-qubit quantum logic gates on \Ca{43} hyperfine qubits with a fidelity $> 99\%$, showing that all quantum logic operations can be performed using this ion with a precision at or above the current state-of-the-art~\cite{Benhelm08,Kirchmair09}.



\section{SUPPLEMENTARY INFORMATION}

\section{Microwave-enhanced optical pumping}

Qubit state preparation begins with optical pumping to the \hfslev{S}{1/2}{4,+4} state using circularly ($\sigma^{+}$) polarized 397\nm\ light, containing two frequencies close to resonance with the $\hfslev{S}{1/2}{3}\leftrightarrow\hfslev{P}{1/2}{4}$ and $\hfslev{S}{1/2}{4}\leftrightarrow\hfslev{P}{1/2}{4}$ transitions. The fidelity of the optical pumping is limited by imperfect polarization, leaving population in (predominantly) ground level $M=+3$ and $M=+2$ states. We use a microwave technique to reduce this error: after clearing out the \hfslev{S}{1/2}{3} states with a pulse of single-frequency 397\nm\ $\sigma^+$ light on the $\hfslev{S}{1/2}{3}\leftrightarrow\hfslev{P}{1/2}{4}$ transition, we apply microwave $\pi$-pulses on the $\hfslev{S}{1/2}{4,+3}\rightarrow\hfslev{S}{1/2}{3,+3}$ and $\hfslev{S}{1/2}{4,+2}\rightarrow\hfslev{S}{1/2}{3,+2}$ transitions, followed by another 397\nm\ $\sigma^+$ pulse to clear out \hfslev{S}{1/2}{3}. The sequence of microwave and clear-out pulses can be repeated as often as necessary. From the SPAM experiments described in the following section, we estimate that this sequence prepares the \hfslev{S}{1/2}{4,+4} state with $<1\times 10^{-4}$ error. The robustness of this technique against polarization imperfections may be especially useful in microfabricated traps incorporating miniaturized integrated optics.

\section{State preparation and measurement error contributions}

In the paper we describe an experiment to determine the combined state preparation and measurement (SPAM) error, consisting of the following steps: (1) microwave-enhanced optical pumping to the \hfslev{S}{1/2}{4,+4} state; (2) three (or four) microwave $\pi$-pulses to transfer the ion to the \ket{\uparrow} (or \ket{\downarrow}) qubit state; (3) four microwave $\pi$-pulses which transfer population from \ket{\uparrow} to $\hfslev{S}{1/2}{4,+4}$ and from \ket{\downarrow} to $\hfslev{S}{1/2}{3,+1}$; (4) ``shelving'' optical pumping pulses to transfer population from $\hfslev{S}{1/2}{4, +4}$ to the metastable 3$\lev{D}{5/2}$ level; (5) time-resolved fluorescence detection to determine whether the ion is shelved. Table I gives a breakdown of the measured SPAM error into contributions from each of these operations. 

The fluorescence detection error (5) was calculated by Monte Carlo simulations using the measured ion fluorescence and background signals, which in our previous experiments using \Ca{40} showed agreement at the $10^{-5}$ level~\cite{Myerson08}. Tests with \Ca{43} in this apparatus showed similar agreement. Drift in experimental conditions (\eg\ the fluorescence signal) increase this error, so the quoted value is a lower bound, but we estimate such effects are at or below the $10^{-5}$ level.

The shelving error (4) was estimated from a rate equation simulation~\cite{DJSthesis}. In this simulation~\citesec{TPHthesis}{5.3} we assumed that all laser detunings, intensities and polarizations were set to their intended values, with the exception of the 393\nm\ intensity. We calibrated the 393\nm\ intensity by measuring the rates at which population was pumped out of the \hfslev{S}{1/2}{4,+4} and \ket{\uparrow} states by the 393\nm\ laser. We found that the pumping rate from \ket{\uparrow} was 15\% greater than that expected from the pumping rate measured for \hfslev{S}{1/2}{4,+4}. (This discrepancy could be explained by a broadening of the 393\nm\ transition, for example due to thermal motion of the ion, or by spectral impurities in the laser.) We used these measured rates in the shelving simulation, resulting in an estimated error somewhat higher than the minimum theoretically attainable. 

To place a limit on the \hfslev{S}{1/2}{4,+4} preparation error (1), we also measured the error in preparing and reading out this state. This experiment was equivalent to the \ket{\uparrow} qubit state measurement with the microwave pulses (2) and (3) removed. We measured an error of $3.6(5)\times 10^{-4}$, which is consistent with the sum of the errors calculated for this state for the shelving transfer ($1.0\times10^{-4}$) and the fluorescence detection ($2.9\times10^{-4}$). This result suggests firstly that the shelving and fluorescence detection errors were not significantly greater than their simulated values; and secondly that the error in preparing the \hfslev{S}{1/2}{4,+4} state was comparable with or smaller than the measurement's uncertainty. (Due to the large Zeeman splittings introduced by the 146\G\ magnetic field, all ground level states other than \hfslev{S}{1/2}{4,+4} have a probability of $\ltish 30\%$ of being shelved~\citesec{TPHthesis}{5.3}; we thus measure the population in the \hfslev{S}{1/2}{4,+4} state, rather than the total population in the $F=4$ manifold, as would be the case at low magnetic fields.) We take $1\times 10^{-4}$ as a conservative upper limit for the \hfslev{S}{1/2}{4,+4} preparation error. 

We attribute the remainder of the measured SPAM error to the microwave $\pi$-pulses, apportioning it equally between the microwave preparation pulses (2) and the microwave readout pulses (3). The average error of $\ish 0.5\,\times10^{-4}$ per pulse is consistent with numerical simulations taking into account off-resonant excitation of spectator transitions and the effect of magnetic field noise ($\ish 1\mG$) on the field-sensitive transitions~\citesec{TPHthesis}{5.2}.

\section{Microwave drive system}

We describe briefly the phase-agile microwave drive system used for the single-qubit randomized benchmarking experiments; further details are given in~\citesec{TPHthesis}{5.7}. 

The microwaves were generated by a frequency-octupled direct digital synthesizer (DDS) running at 400\MHz, referenced to a rubidium frequency standard. The DDS was programmed with four different phases, so as to generate microwaves with four 90\degree-separated phases to drive $\pi/2$ rotations about $\pm x$ and $\pm y$ axes; the phase was switched during the dead time (14\us) between $\pi/2$-pulses. The DDS and octupler were run continuously and at a constant power level to minimize effects of experimental duty cycle. A pair of high-isolation switches in series was used to produce nominally rectangular pulses, which were fed to one of the trap's axial electrodes via a circulator and filters as indicated in figure~1a in the paper. The nominal extinction of the switches was 60\dB\ each, and the microwave frequency was detuned by $+4.5\Hz$ from the qubit transition which further reduced the effects of any leakage. The DDS amplitude was set to produce a $\pi/2$-pulse duration of 12.1\us\ and periodically calibrated using a sequence of 751 $\pi/2$-pulses. At the input to the trap, the microwave power was approximately 0.1\mW.

\section{Single-qubit gate error contributions}

In this section we consider the three sources of error in the single-qubit randomized benchmarking experiment listed in Table I: (1) microwave detuning from the qubit transition; (2) incorrect microwave $\pi/2$-pulse area; (3) off-resonant effects due to spectator transitions in the ground level. A more detailed error analysis for this experiment may be found in~\citesec{TPHthesis}{5.7}, which considers the following additional sources of error: qubit dephasing, microwave phase and amplitude noise, pulse length jitter, microwave-ion coupling fluctuations, microwave and laser leakage, ion heating and effects of microwave duty cycle. These are all estimated to be insignificant compared with the errors considered here.

Our method is as follows. We simulate the error for a given sequence of gates by calculating a propagator describing each (imperfect) $\pi/2$-pulse and 14\us\ inter-pulse dead time in the sequence. These are multiplied together to create a sequence propagator, from which the expected error for that sequence may be calculated. By averaging over the particular sequences used in the benchmarking experiment, we calculate the expected average error per gate (EPG). For simplicity, we consider only the longest sequences used (2000 computational gates), as the final result is most sensitive to these runs. To verify the simulations, we performed additional randomized benchmarking experiments in which we introduced intentional microwave detuning and pulse area errors. 

\begin{figure}
\includegraphics[width=\onecolfig]{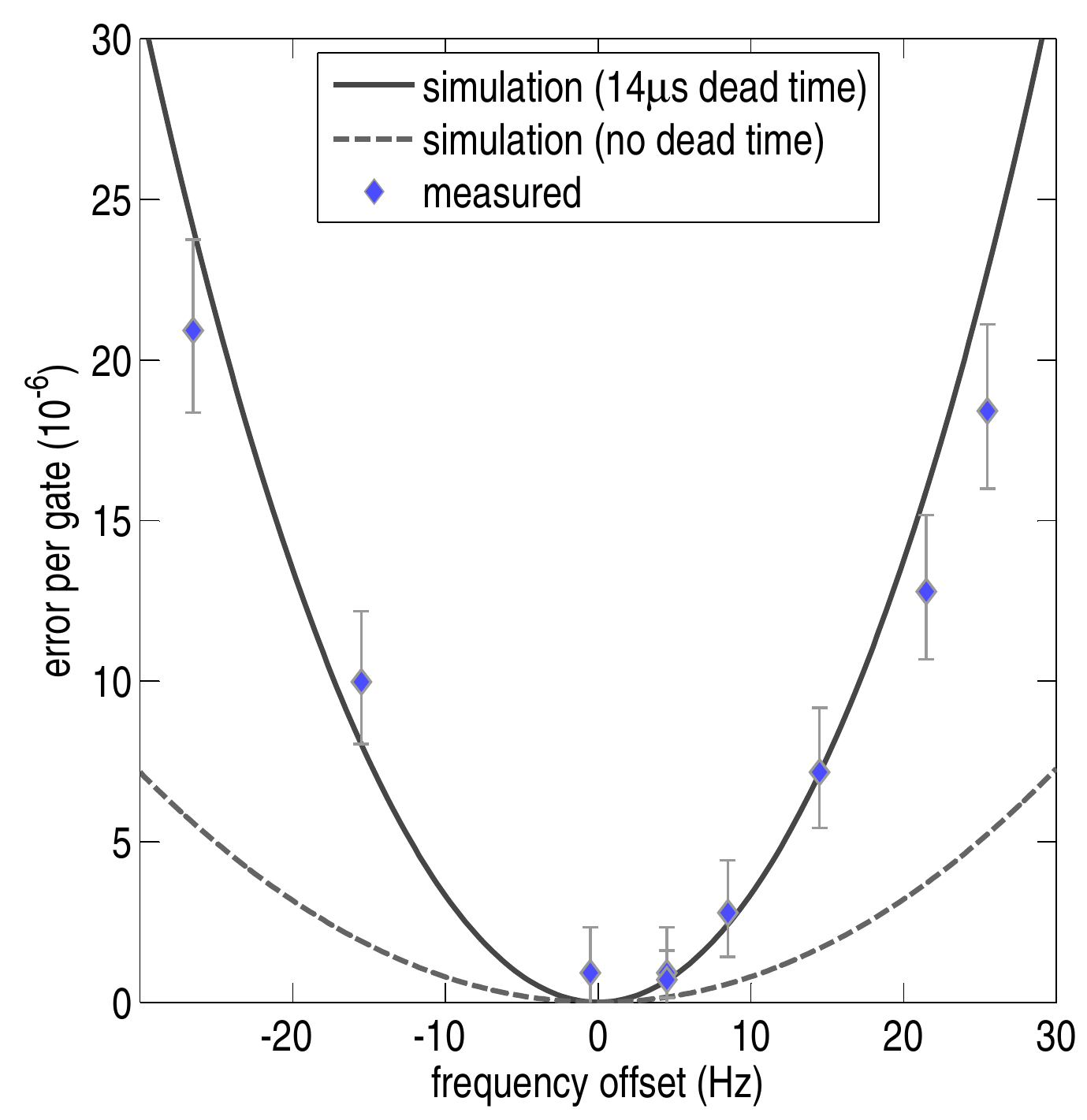}
\caption{%
Simulated and measured EPG as a function of microwave detuning from the qubit transition. We simulate errors both with and without the 14\us\ inter-pulse dead time. These measurements used 2000-gate sequences and have been corrected for SPAM error. The experiments reported in the paper were performed with a detuning of $+4.5\Hz$. Including the effect of the a.c.\ Zeeman shift makes negligible difference to the simulated EPG on this scale.
}
\label{F:freqErr}
\end{figure}

Figure~\ref{F:freqErr} shows the simulated and measured EPG as a function of the microwave detuning from the qubit transition. The measured results are in good agreement with the simulations. We performed simulations both with and without experimental dead time, demonstrating that the dead time makes the experiment significantly more sensitive to detuning error. Due to the presence of off-resonant spectator transitions which couple to the qubit states, there is an a.c.\ Zeeman shift of the qubit transition which we calculate to be $-1.0\Hz$ at our Rabi frequency of $\Omega/2\pi = 21\kHz$. For the experiments reported in the paper, the  microwave detuning was $+4.5\Hz$; the a.c.\ Zeeman shift increases this to $+5.5\Hz$ during the $\pi/2$-pulses. Including this effect, the simulated EPG is $0.7\times 10^{-6}$.

\begin{figure}
\includegraphics[width=\onecolfig]{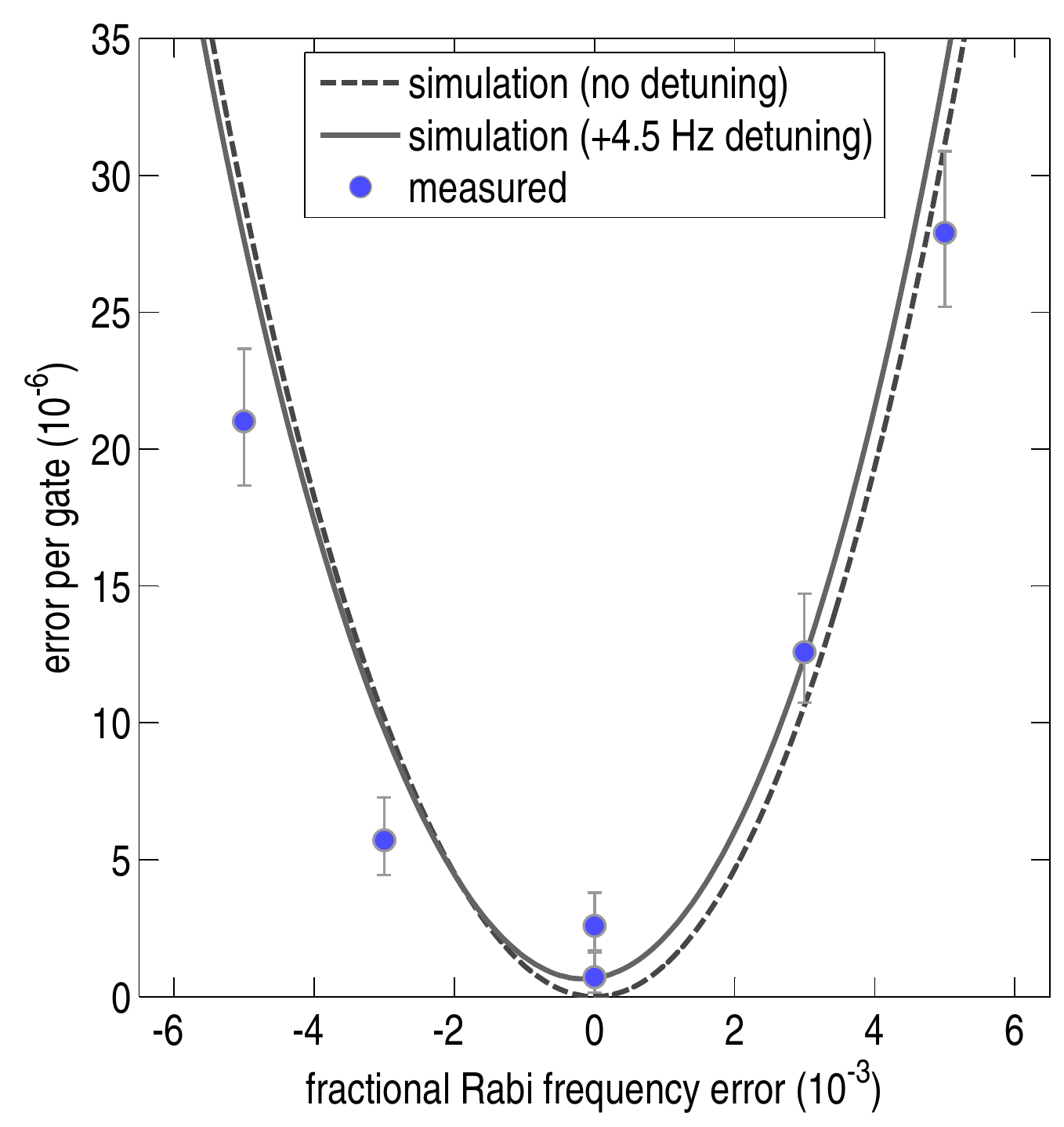}
\caption{%
Simulated and measured EPG as a function of microwave $\pi/2$-pulse area error (varied by changing the Rabi frequency, while keeping the pulse duration fixed). We simulate both microwaves resonant with the qubit transition (dashed curve) and microwaves with the $+4.5$\Hz\ detuning used in the main experiment (solid curve). These measurements were also taken with a $+4.5$\Hz\ detuning, used 2000-gate sequences, and have been corrected for SPAM error.
}
\label{F:ampErr}
\end{figure}

Figure~\ref{F:ampErr} shows the simulated and measured EPG as a function of Rabi frequency for fixed pulse duration. Again, we find good agreement between measurement and simulation. For the experiments reported in the paper, we periodically recalibrated the microwave amplitude during the randomized benchmarking experiments. Based on the required changes to the DDS amplitude at each recalibration, we estimate that the microwave Rabi frequency drifted by $\ltish 5\times10^{-4}$ between calibrations. We attribute this to slow drift due to thermal effects in the microwave system, which we model as a constant Rabi frequency offset, giving an EPG contribution of $0.3\times 10^{-6}$.

Since the microwave field used to drive the qubit transition was not purely $\sigma^+$ polarized, it also couples off-resonantly to other ground-level transitions. In addition to the a.c.\ Zeeman shift of the qubit transition discussed above, this also leads to Rabi flopping on these spectator transitions with a small amplitude $\ish(\Omega/\Delta\sub{Z})^2\,\ish 10^{-7}$ for the Zeeman splittings $\Delta\sub{Z}/2\pi \approx 50\MHz$ at 146\G. To calculate the error due to off-resonant Rabi-flopping, we simulated the randomized benchmarking experiment including the full 16-state ground level and the known microwave polarization (measured in separate experiments). This simulation gives an error $0.1\times 10^{-6}$, confirming the rough estimate above.

\section{Random sequence sample size}

\begin{figure}
\includegraphics[width=\onecolfig]{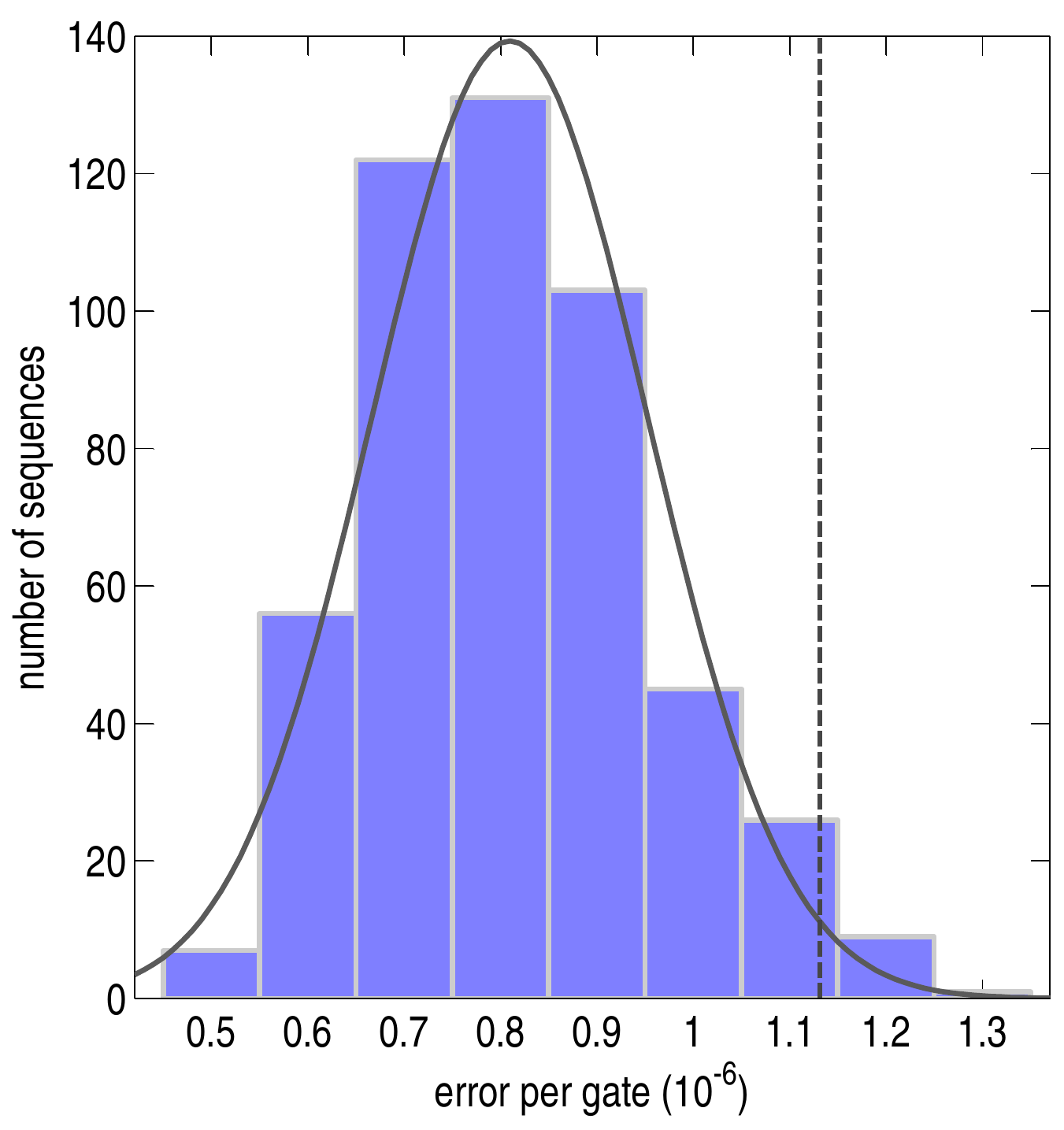}
\caption{%
Histogram of the simulated EPG from 500 randomly-chosen different sets of 32 sequences, each of length 2000 computational gates. The simulation used 12.1\us\ pulses with 14\us\ inter-pulse dead-times, a detuning of $+4.5\Hz$ and a constant Rabi frequency error of $5\times10^{-4}$. The black curve is the best fit normal distribution, with mean $\mu=0.81\times10^{-6}$ and standard deviation $\sigma=0.14\times10^{-6}$. The dashed line shows the simulated EPG for the particular set of sequences used in our experiment.
}
\label{F:RBMsampling}
\end{figure}

Randomized benchmarking aims to determine the mean EPG averaged over all possible pulse sequences. In practice, of course, only a finite number of randomly-chosen sequences can be tested: in the experiments reported in the paper, we used 32 sequences for each different sequence length. To confirm that this was a sufficient level of randomization to provide an accurate estimate of the mean EPG, we performed our simulation of the experiment on 500 different sets of 32 random sequences, each consisting of 2000 computational gates. These simulations assumed the nominal conditions of the experiment (neglecting off-resonant effects). A histogram of the simulated EPGs is shown in figure~\ref{F:RBMsampling}: the standard deviation of the results is a factor of two smaller than the uncertainty in our measured EPG, implying that this level of randomization is sufficient. The particular set of sequences we used in our experiment lies about $2\sigma$ above the mean, suggesting that it has above-average sensitivity to known experimental imperfections.
\vfill

\section{Acknowledgements}

It is a pleasure to thank E. Knill and A. Meier for advice regarding the randomized benchmarking experiment, A. Steane and J. Jones for helpful discussions, and J. Home for comments on the manuscript. This work was supported by the EPSRC.



\newpage


\end{document}